\documentclass[prx,twocolumn,floatfix,superscriptaddress]{revtex4-2}

\usepackage{graphicx}
\usepackage{color}
\usepackage[normalem]{ulem}
\usepackage{braket}
\usepackage{amsmath}
\usepackage{verbatim}
\usepackage{amssymb}

\newcommand{\be}{\begin{equation}}
\newcommand{\ee}{\end{equation}}
\newcommand{\ba}{\begin{eqnarray}}
\newcommand{\ea}{\end{eqnarray}}
\newcommand{\nn}{\nonumber \\}

\begin{document}

 \title{Quantum error mitigation in quantum annealing}

\newcommand{\affildw}{D-Wave Systems, Burnaby, British Columbia, Canada}
\newcommand{\affilsfu}{Department of Physics, Simon Fraser University, Burnaby, British Columbia, Canada}
\newcommand{\affilubc}{Department of Physics and Astronomy and Quantum Matter
  Institute, The University of British Columbia, Vancouver, British Columbia, Canada}

\author{Mohammad H. Amin}
\email[]{amin@dwavesys.com}
\affiliation{\affildw}
\affiliation{\affilsfu}

\author{Andrew D. King}
\affiliation{\affildw}

\author{Jack Raymond}
\affiliation{\affildw}

\author{Richard Harris}
\affiliation{\affildw}

\author{William Bernoudy}
\affiliation{\affildw}

\author{Andrew J. Berkley}
\affiliation{\affildw}

\author{Kelly Boothby}
\affiliation{\affildw}

\author{Anatoly Smirnov}
\affiliation{\affildw}

\author{Fabio Altomare}
\affiliation{\affildw}

\author{Michael Babcock}
\affiliation{\affildw} 

\author{Catia Baron}
\affiliation{\affildw}

\author{Jake Connor}
\affiliation{\affildw}




\author{Martin Dehn}
\affiliation{\affildw}



\author{Colin Enderud}
\affiliation{\affildw}

\author{Emile Hoskinson}
\affiliation{\affildw}

\author{Shuiyuan Huang}
\affiliation{\affildw}

\author{Mark W. Johnson}
\affiliation{\affildw}

\author{Eric Ladizinsky}
\affiliation{\affildw}

\author{Trevor Lanting}
\affiliation{\affildw}

\author{Allison J. R. MacDonald}
\affiliation{\affildw}

\author{Gaelen Marsden}
\affiliation{\affildw}

\author{Reza Molavi}
\affiliation{\affildw}



\author{Travis Oh}
\affiliation{\affildw}






\author{Gabriel Poulin-Lamarre}
\affiliation{\affildw}


\author{Hugh Ramp}
\affiliation{\affildw}

\author{Chris Rich}
\affiliation{\affildw}



\author{Berta Trullas Clavera}
\affiliation{\affildw}

\author{Nicholas Tsai}
\affiliation{\affildw}

\author{Mark Volkmann}
\affiliation{\affildw}


\author{Jed D. Whittaker}
\affiliation{\affildw}

\author{Jason Yao}
\affiliation{\affildw}

\author{Niclas Heinsdorf}
\affiliation{\affilubc}

\author{Nitin Kaushal}
\affiliation{\affilubc}

\author{Alberto Nocera}
\affiliation{\affilubc}

\author{Marcel Franz}
\affiliation{\affilubc}

\begin{abstract}

Quantum Error Mitigation (QEM) presents a promising near-term approach to reduce error when estimating expectation values in quantum computing. Here, we introduce QEM techniques tailored for quantum annealing, using Zero-Noise Extrapolation (ZNE). We implement ZNE through zero-temperature extrapolation as well as energy-time rescaling. We conduct experimental investigations into the quantum critical dynamics of a transverse-field Ising spin chain, demonstrating the successful mitigation of thermal noise through both of these techniques. Moreover, we show that energy-time rescaling effectively mitigates control errors in the coherent regime where the effect of thermal noise is minimal. Our ZNE results agree with exact calculations of the coherent evolution over a range of annealing times that exceeds the coherent annealing range by almost an order of magnitude.

\end{abstract}

 \maketitle

\section{Introduction}

Quantum computation has emerged as a computational paradigm with the potential to solve complex problems that remain beyond the capabilities of classical computers. Nevertheless, its performance is substantially hampered by environmental noise and hardware imperfections. Quantum Error Correction (QEC) \cite{Shor95,Steane96} is regarded as the ultimate solution to eliminate the impact of these errors. However, the significant overhead of QEC limits its practicality only to very large scales, far beyond the current state of technology \cite{Preskill18}. Recently, Quantum Error Mitigation (QEM) has been proposed \cite{Temme17,Li17,Endo18,Cai22} as a near-term solution that can be used to estimate error-free expectation values when the impact of noise is small. Among the various QEM techniques, Zero-Noise Extrapolation (ZNE) \cite{Temme17,Li17} stands out as one of the most practical methods.  In ZNE, one systematically varies the noise amplitude experienced by the system. By observing the system's response to this controlled change, it becomes possible to make predictions about how the system would behave under noise-free conditions. 

While QEM was initially developed and tested for circuit model quantum computing \cite{Kandala19,PEC,Kim23a,Kim23,Ritajit23}, the same principles can be applied to other quantum computing protocols, including quantum annealing (QA) \cite{Johnson11}. Quantum Anneling Correction (QAC), founded on repetition codes, has been proposed and examined as a means to enhance the performance of QA at the expense of requiring a larger number of qubits \cite{QAC1,QAC2,QAC3,QAC4,QAC5,QAC6,QAC7,QAC8,QAC9,QAC10,QAC11}. Over the past few years, quantum simulation has evolved into an important application of QA for the exploration of exotic states of condensed matter systems \cite{Harris18, King18,SpinIce,KagomeIce} and their quantum phase transitions \cite{King22, King23}.  QEM can enhance the accuracy of expectation values in quantum simulation experiments without incurring any additional overhead in qubit count. Indeed, the first experimental attempt to extract information about noise-free evolution via extrapolation was reported in Ref.~\cite{Dickson13}, in which QA performance was examined as a function of temperature for a problem with a very small spectral gap. By extrapolating the final ground state probability to zero temperature, the noise-free Landau-Zener behavior was replicated (see Fig. 3 and 4 of Ref.~\cite{Dickson13}).

In this paper, we provide a theoretical description of ZNE in quantum annealing, with a particular emphasis on extracting information about quantum phase transitions. We then employ ZNE in experimental investigations into the critical dynamics of a 1D quantum Ising spin chain.  The extrapolated results compare well with exact solutions as well as time dependent density matrix renormalization group (DMRG) simulations for the closed system Schr\"{o}dinger evolution across a wide range of control parameters.

\section{Theory}

The Hamiltonian of an annealing-based quantum processing unit (QPU) is written as
\begin{equation}\label{HS}
  H_S(s) = -\Gamma(s)\sum_i \sigma_i^x  + \mathcal J(s)  H_P,
\end{equation}
where
\be \label{HP}
H_P = \sum_i h_i \sigma_i^z + \sum_{\langle i,j\rangle} J_{ij} \sigma_i^z\sigma_j^z
\ee
is the problem Hamiltonian and $s=t/t_a$, with $t$ being time and $t_a$ being the annealing time. The energy scales $\Gamma(s)$ and $\mathcal J(s)$ evolve with time in such a way that $\Gamma(0) \gg \mathcal J(0)$ and $\Gamma(1) \ll \mathcal J(1)$. They are predetermined at the time of calibration (see Fig.~\ref{fig:schedule} in Appendix C). The dimensionless biases ($h_i$) and coupling coefficients ($J_{ij}$) are programmable.

To incorporate all sources of error, whether thermal, $1/f$ noise, or parameter misspecification, we consider the qubits to be in contact with their environment. The total Hamiltonian is written as
\be
H = H_S + H_B + H_{\rm int}, \label{HSB}
\ee
where $H_B$ is the environment Hamiltonian, and
\be
 H_{\rm int} = - \sqrt{\gamma} \sum_{\alpha} Q^\alpha {\cal O}^\alpha \label{Hint}
\ee
describes the interaction between system and environment. The operators ${\cal O}^\alpha$ are local operators acting on the qubits. They are typically Pauli matrices (e.g., $\sigma^z_i$ for flux noise and $\sigma^y_i$ for charge noise), but such details are irrelevant for the purpose of this work. The operators $Q^\alpha$ act on the corresponding environments, with $\alpha$ spanning all sources of noise. We include control error within the same framework by taking $Q^\alpha$ to be a constant random number. For example, to incorporate static errors $\delta h_i$ ($\delta J_{ij}$) in the Hamiltonian parameters, we use ${\cal O}^\alpha = \sigma^z_i \ (\sigma^z_i\sigma^z_j)$ and $\sqrt{\gamma} Q^\alpha = \delta h_i \ (\delta J_{ij})$. The corresponding noise spectral density in that case is a $\delta$-function at zero frequency. The prefactor $\sqrt{\gamma}$ is introduced to keep track of the perturbation order; $\gamma$ can be thought of as an overall decay rate, in units of inverse time, multiplying all relaxation and dephasing rates. Its precise definition is unimportant for our discussion.

The state of an open quantum system is commonly described by the reduced density matrix $\rho(t)$, which is the solution to a master equation, e.g., Bloch-Redfield \cite{BR} or Lindblad \cite{Lindblad,LidarLambShift}. As we show in Appendix A, the density matrix at the end of the evolution can be expanded in powers of $\gamma$ as
\be \label{rhoexpansion}
\rho(t_a) =  \sum_{\mu=0}^\infty (\gamma t_a)^\mu \rho_\mu(t_a)
\ee
where $\rho_0(t_a)$ is the noise-free contribution to the density matrix and $\rho_{\mu>0} (t_a)$ take care of perturbative corrections due to the environment. As such, any observable $A$ at the end of annealing can be expanded as
\be \label{Ata}
\langle A(t_a) \rangle = \sum_{\mu = 0}^\infty \langle A(t_a) \rangle_\mu,
\ee
where
\be \label{Atalpha}
\langle A(t_a) \rangle_\mu = \text{Tr} [(\gamma t_a)^\mu \rho_\mu(t_a) A].
\ee
The zeroth order term, $\langle A(t_a)\rangle_0$, is the error-free expectation value for coherent evolution. Our goal is to extract $\langle A(t_a)\rangle_0$ by extrapolation from a set of noisy evolutions.  

ZNE is implemented by amplifying the effect of noise by a controllable factor $\lambda$ in such a way that $\rho_0(t_a)$ and therefore $\langle A(t_a)\rangle_0$ remain unchanged. This can be easily achieved by increasing the coupling coefficient:
\be 
\gamma \to \lambda \gamma.
\label{gamma_scaling}
\ee
Substituting \eqref{gamma_scaling} into \eqref{Atalpha} and \eqref{Ata} and expanding up to order $M$, we obtain
\be \label{Aexpand}
\langle A(t_a,\lambda) \rangle \approx  \sum_{\mu = 0}^M C_\mu \lambda^\mu ,
\ee
where $C_\mu = \langle A(t_a) \rangle_\mu$. We can now estimate the error-free expectation value, $\langle A(t_a) \rangle_0 = C_0$, by measuring $\langle A(t_a,\lambda) \rangle$ for a range of $\lambda$-values, fitting the result into \eqref{Aexpand}, and extrapolating back to $\lambda \to 0$. When the effect of the environment is small, only a few terms in the expansion (usually $M=1$ or 2) suffice to yield a good estimate. In practice, it is not feasible to amplify noise by increasing $\gamma$ because it requires manipulating the qubits' environment at the microscopic level. However, it is possible to mimic such an effect by other means, as we shall discuss next. 

\section{Noise amplification by temperature}

The simplest way to increase the influence of a thermal environment is by increasing its temperature $T$. Clearly, the coherent contribution to the density matrix $\rho_0(t_a)$ remains not affected by changing $T$. As we show in Appendix A, in the regime of weak coupling to the environment, all decay processes, including relaxation and dephasing, are functions of the thermal noise spectral density, which we denote by $S(\omega)$ \cite{Note}. For example, the relaxation rate between states $\ket{n}$ and $\ket{m}$ is proportional to $S(\omega_{nm})$, where $\omega_{nm} = E_n - E_m$ is the energy difference between the two states, and the dephasing rates are functions of $S(0)$ or spectral density at low frequencies. For an ohmic environment in thermal equilibrium at temperature $T$, the noise spectral density is given by ($\hbar = k_B = 1$)
\be
S (\omega) = {\eta \omega \over 1- e^{-\omega/T}},
\ee
where $\eta$ is a constant. When the evolution of the system is slow (near adiabatic) only low energy eigenstates of the instantaneous Hamiltonian get occupied. Thus, only transitions between these states, for which $\omega_{nm} < T$, affect the density matrix. We can therefore expand $e^{-\omega/T}$ in the denominator to first order to obtain $S (\omega) \approx \eta T$.
This means that 
\be
T \to \lambda T \quad \Longrightarrow \quad S (\omega) \to \lambda S(\omega).
\ee
Since the coefficient $\lambda$ multiplies all decay rates, it can be factored out and absorbed into the coupling coefficient $\gamma$, leading to \eqref{gamma_scaling}. Factoring out $\lambda$ would keep $\rho_\mu(t_a)$ unchanged in \eqref{rhoexpansion}, resulting in expansion \eqref{Aexpand}, thus allowing ZNE. 

While the above explanation relied on low energy expansion of an Ohmic spectral density, the conclusion goes beyond such assumptions. As long as the change in $T$ can be treated as a perturbation, an expansion similar to \eqref{Aexpand} is expected, but with different coefficients $C_\mu$ and  with $C_0$ representing the zero-temperature expectation value, which is not necessarily equivalent the error-free one (see below). 

\begin{figure}
\begin{center}
\includegraphics[scale=0.8]{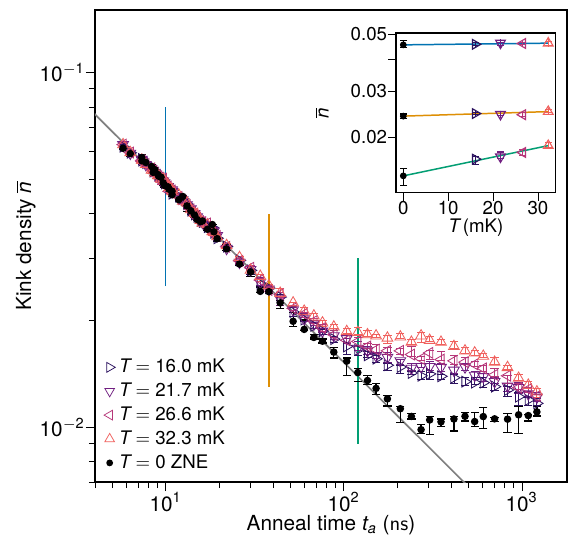}
\caption{\label{fig1} Error mitigation by temperature extrapolation. Main panel: Kink density as a function of annealing time for four different temperatures. The gray straight line represents the theoretical prediction for error-free coherent dynamics $n\propto t_a^{-1/2}$. The $T$-dependent deviation from the prediction is due to thermal excitations. The solid black circles represent ZNE results at each $t_a$. The extrapolated results follow the theoretical prediction for an extended range of $t_a$ compared to unmitigated data. Inset: extrapolations in $T$ for three annealing times $t_a$ indicated by the three color-coded vertical lines in the main plot. }
\end{center}
\end{figure}

We experimentally tested zero-$T$ extrapolation using a prototype D-Wave Advantage2$^{\rm TM}$ processor, with 232 flux qubits coupled in a one-dimensional periodic chain. We applied zero bias $h_i=0$ to all qubits and used uniform ferromagnetic coupling $J_{ij} = -1.8$.  We annealed the QPU using fast anneal to simulate critical dynamics of a transverse-field Ising spin chain. Details of the experiment are provided in Appendix C and Ref.~\cite{King22}. 
We measured kink density defined as
\be
n = {1\over L} \sum_{i=1}^L \langle K_i \rangle
\ee
where $L$ is the length of the chain and
\be
K_i = (1-\sigma^z_i\sigma^z_{i+1})/2
\label{Ki}
\ee
is the kink operator. Figure~\ref{fig1} shows $n$ as a function of annealing time $t_a$. The triangle symbols represent QPU results obtained at different temperatures and the solid lines represent the theoretical prediction, $n\propto t_a^{-1/2}$, according to Jordan-Wigner transformation and DMRG simulations using the time dependent variational principle (TDVP). (details about the DMRG simulations are provided in Appendix D). The experimental data follow the theoretical predictions up to $t_a\lesssim 40\,$ns, above which thermal excitations generate additional kinks \cite{King22,Bando2020}. The inset shows data corresponding to the three vertical cross sections in the main panel, distinguished by three colors in both plots. Linear extrapolations to $T=0$ for these three cross sections are indicated by black circles in both panels. The extrapolated points follow the power law $n\propto t_a^{-1/2}$, as predicted by coherent theory, up to $t_a\lesssim200\,$ns and then deviate sharply from the theoretical prediction. Beyond this point thermal excitations cannot be treated as perturbation.

While zero temperature extrapolation seems to reproduce correct results for the kink density, it is not expected to correct miscalibration or other nonthermal errors that do not depend on $T$. These errors can affect other physical observables such as higher order correlations, as we shall discuss later. Moreover, our QPU architecture currently does not support rapid changes in temperature and therefore ZNE in $T$ cannot be regarded as an efficient way of improving simulation accuracy.

\section{Noise amplification by energy-time rescaling}

\begin{figure*}
\begin{center}
\includegraphics[scale=0.75]{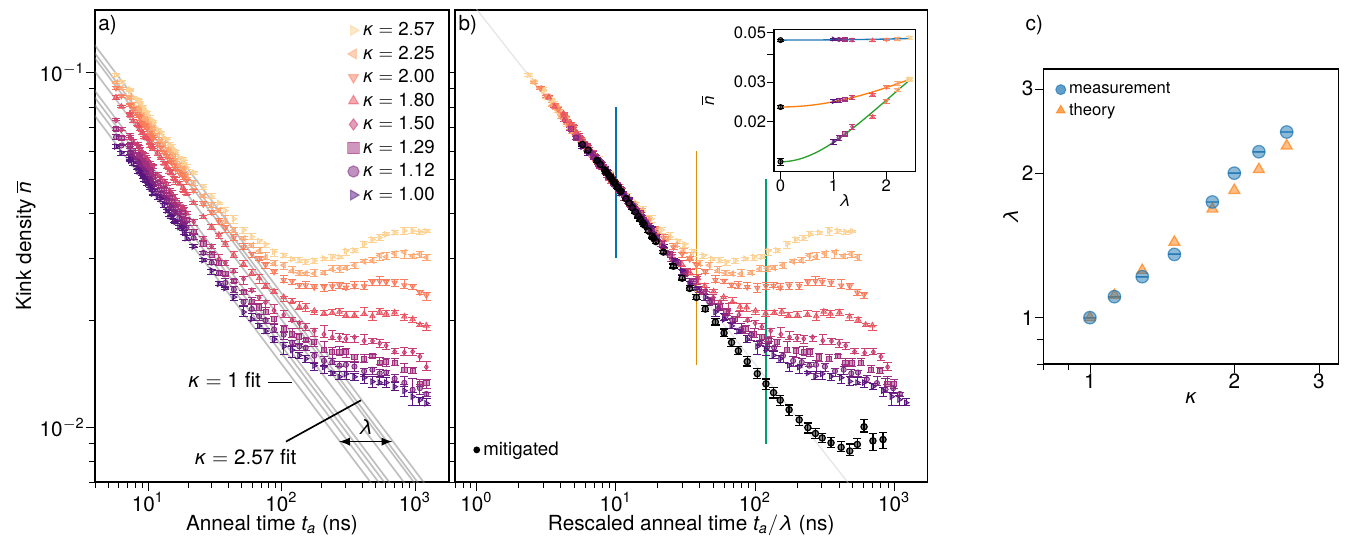}
\caption{\label{fig2} Error mitigation by energy extrapolation. (a) Kink density is measured for various annealing times $t_a$ and coupling energies $J=-1.8/\kappa$, with $\kappa$ listed in the legend of panel b.  An empirical rescaling of time is performed by collapsing fits of $\rho \propto t_a^{-1/2}$. (b) Collapse of panel a data plotted with rescaled x-axis. Circles represent the extrapolated points. Inset: ZNE is performed in $\lambda$, using a quadratic fit, Eq.~\eqref{QZNE}. (c) $\lambda(\kappa)$ obtained from collapse (circles) compared to those obtained from Eq.~\eqref{lambda-kappa} (triangles). }
\end{center}
\end{figure*}

A more effective way of amplifying noise is via energy-time rescaling. Since $\gamma$ is multiplied by $t_a$ in \eqref{rhoexpansion}, it is possible to mimic the effect of increasing $\gamma$ by increasing $t_a$. However, in order to keep the noiseless unitary evolution $\rho_0(t_a)$ from being affected, one needs to rescale both anneal time and energy. This can be achieved by substituting $t_a \to \lambda t_a$ and $H_S \to H_S/\lambda$, which keeps $\int_0^{t_a} H_S(t) dt$ and therefore the unitary time evolution operator unchanged. In gate model quantum computation, this approach is referred to as analog-ZNE, in contrast to digital-ZNE, in which noise is amplified by inserting additional (identity) gates \cite{Ritajit23}. Intuitively, extending the annealing time would increase the effect of relaxation and dephasing on the qubits.  Reducing the energy scale, on the other hand, would increase the relative importance of control errors as long as they remain unchanged by the rescaling. These all lead to a  $\lambda$-dependent reduction of accuracy. It should be noted that $\rho_{\mu>0}(t_a)$ may also have some residual dependence on $\lambda$, but that can be easily incorporated into the Taylor expansion \eqref{Aexpand} by redefining the expansion coefficients $C_\mu$. This means that only the invariance of $\rho_0(t_a)$ is required for ZNE to work.

In QA processors, rescaling the system Hamiltonian $H_S$ in \eqref{HS} is challenging because both $\Gamma(s)$ and ${\cal J}(s)$ are fixed functions. However, one can easily scale down $H_P$ by reprogramming $h_i$ and $J_{ij}$ in \eqref{HP}. Since only one part of $H_S$ is now rescaled, making sure the coherent evolution remains unaffected is nontrivial. Let us introduce an energy-time rescaling as
\be
t_a \to \lambda t_a, \qquad 
H_P \to H_P/\kappa .
\ee
We need to choose $\lambda$ and $\kappa$ in such a way that $\rho_0(t_a)$ remains unchanged. In general, this may not be exactly possible, but can be achieved approximately. When the evolution is dominated by a quantum phase transition, one can obtain an approximate $\lambda(\kappa)$ in such a way that the critical dynamics remains unaffected by the rescaling (see Appendix B for details). This is possible because critical dynamics occur within a narrow region close to the critical point $s_c$ and the statistics only depend on the speed of passing this point. Rescaling $H_P$ would move the critical point to a new point $s_c^\kappa \equiv s_c(\kappa)$ with a new effective speed. This change has to be compensated by rescaling the annealing time. For the case of the transverse-field Ising spin chain studied here, one can show that (see Appendix B)
\be
\lambda(\kappa) = { \mathcal J(s_c)^2 [ \mathcal J'(s_c^\kappa) - \Gamma'(s_c^\kappa)] \over \mathcal J(s_c^\kappa)^{2} [ \mathcal J'(s_c) - \Gamma'(s_c) ]}.
\label{lambda-kappa}
\ee
where primes indicate first derivatives with respect to $s$.  It is also possible to determine $\lambda(\kappa)$ experimentally as we discuss below. Once $\lambda$ is determined, we can use \eqref{Aexpand} to fit the data and extrapolate to $\lambda = 0$. It should be emphasized that $\lambda \to 0$ corresponds to very short annealing times \cite{Note2}.

\begin{figure*}
\begin{center}
\includegraphics[scale=0.8]{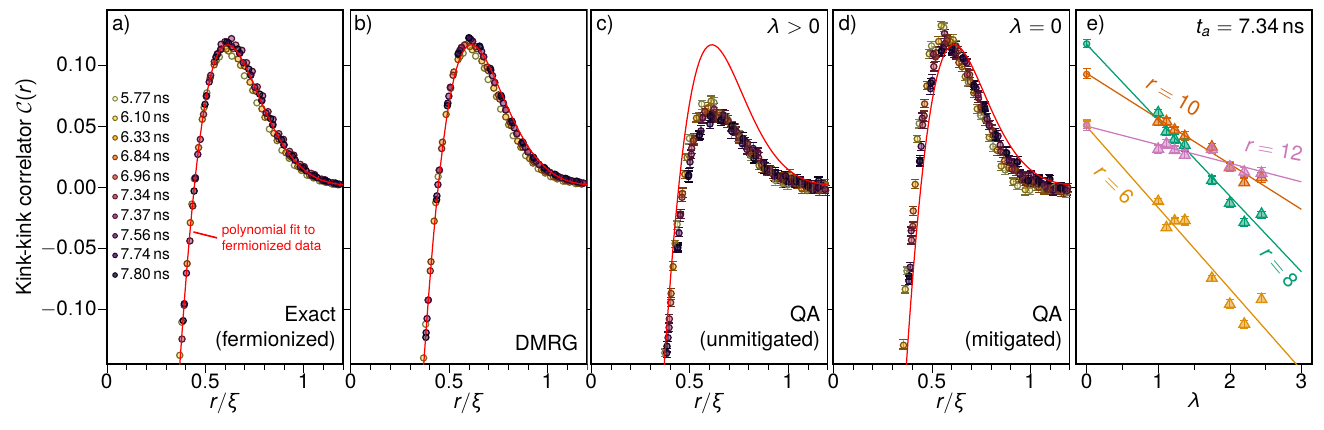}
\caption{\label{fig3} Mitigation of kink-kink correlation by energy extrapolation.   We compute the observable ${\cal C}(r)$ by  (a) fermionizing the system via Jordan-Wigner transformation, and (b) using numerics with time dependent DMRG using the TDVP algorithm (see Appendix D for details).  Points fall approximately on a non-universal curve for a range of $t_a$.  Red line indicates a degree-10 polynomial that is replicated on the other panels to the right to facilitate comparison between theory and experiment.  (c) Unmitigated QA data show correct qualitative behavior, but with a significantly flattened peak speculated to arise primarily due to unintended Hamiltonian disorder.  (d) Results obtained via linear ZNE to $\lambda=0$ show close agreement with the fermionized solution.  (e) We show example extrapolations for varying lattice distances $r$, with $t_a=7.34$\, ns.}
\end{center}
\end{figure*}

For the case of thermal noise, the environment does not affect the system when the annealing time is shorter than the thermal relaxation time, hence no $\lambda$-dependence. As such, we should expect a flat curve with zero slope at small $\lambda$, i.e., $C_1 = 0$. This means that for thermal noise, the lowest order polynomial is quadratic: 
\be
\langle A(t_a,\lambda) \rangle \approx C_0 + C_2 \lambda^2. \label{QZNE}
\ee

Figure \ref{fig2}a shows measurements of kink density for different values of $J=-1.8/\kappa$. As expected, reducing the energy scale increases the number of kinks. Moreover, it increases the relative influence of the thermal environment as evidenced by the increasingly obvious deviation from $n\propto t_a^{-1/2}$ at high $t_a$.  Nonetheless, it is possible to rescale the horizontal axis $t_a\to t_a/\lambda$, using $\lambda$ as a free parameter, to collapse all of the data within the coherent annealing regime. The resulting collapse is shown in Fig.~\ref{fig2}b and the values of $\lambda(\kappa)$ obtained by this collapse are plotted in Fig.~\ref{fig2}c together with theoretical predictions using \eqref{lambda-kappa}. Note that the collapse method cannot mitigate static errors, such as Hamiltonian misspecification because they affect both coherent and thermally activated evolution. The agreement between the empirical collapse and the annealing schedule inferred theoretical values of $\lambda(\kappa)$ in the inset suggests that kink-density is insensitive to such errors.

The next step involves using extrapolation to obtain a noise-free kink density in the regime where the thermal environment is effective.  Since the dominant noise is thermal, one should extrapolate using quadratic fits per Eq.\ \eqref{QZNE}. The inset in Fig.~\ref{fig2}b illustrates this extrapolation for three different values of $t_a/\lambda$.  The black circles in the main panel show the result of this extrapolation. As in Fig.~\ref{fig1}, the extrapolated points agree with the exact $n\propto t_a^{-1/2}$ behavior for up to $t_a/\lambda\lesssim200\,$ns.  Thus one can successfully infer noise-free expectation values over a much broader range of annealing time than the ostensible coherent annealing regime that is nominally bounded by $t_a\lesssim40\,$ns.

While kink density is unaffected by control error, as the inset of Fig.~\ref{fig2}a suggests, other quantities may not exhibit this resilience. One such quantity is kink-kink correlation \cite{Nowak,Dziarmaga2022}
\be
{\cal C}(r) = {1\over L} \sum_{i=1}^L {\langle K_i K_{i+r} \rangle - n^2 \over n^2} ,
\ee
with $K_i$ defined in Eq.\ \eqref{Ki}. Figure \ref{fig3} shows ${\cal C}(r)$ as a function of $r$ normalized to the correlation length $\xi = 1/n$. Figures~\ref{fig3}a and \ref{fig3}b illustrate results obtained from the exact solution using Jordan-Wigner transformation and DMRG, respectively. The measured ${\cal C}(r)$ depicted in Fig.~\ref{fig3}c exhibits a systematic deviation from the exact results in Fig.~\ref{fig3}a, independent of $t_a$. Such a deviation was first observed in Ref.~\cite{King22} and was attributed to the disorder due to control error or $1/f$ noise. Disorder in the potential energy can attract kinks toward local minima, thus affecting the kink-kink correlation while keeping the density of kinks unchanged. As discussed before, the effect of disorder and control error can be mitigated by ZNE through energy-time rescaling. Figure \ref{fig3}d illustrates the ZNE result, showing a significant improvement in alignment with the exact theoretical predictions (depicted by the solid lines), compared to unmitigated results. In Fig.~\ref{fig3}e, a few examples of extrapolation at different values of r are displayed and, unlike the case of thermal noise (Fig.~\ref{fig2}), linear extrapolation works well.

\section{Conclusion}

We have successfully demonstrated the implementation of Zero-Noise Extrapolation (ZNE) for quantum annealing. Through the analysis of critical dynamics in a 1D quantum spin chain, we mitigated the impact of thermal noise on kink density, extending the effective decoherence time by almost an order of magnitude. Additionally, we mitigated the influence of static errors possibly due to mis-calibration or $1/f$ noise, on coherent evolution, as evidenced by having successfully estimated kink-kink correlations. All extrapolated results are in good agreement with the exact solution to the Schr\"odinger equation obtained from the Jordan-Wigner transformation as well as DMRG. 

It is important to emphasize that ZNE primarily enhances the accuracy of expectation values but does not directly impact sample quality. 
Nevertheless, one can employ extrapolation of probabilities to gain insights into error-free probability distribution. This allows for the reconstruction of the distribution by adjusting sample weights or employing Monte Carlo techniques. For some problems this may require a significant amount of sampling. Extrapolating probabilities is particularly useful when evaluating measures such as time-to-solution \cite{Ronnow} for increasingly lower noise and more coherent systems, hence constructing a performance model for the QPU and providing a means to specify requirements on coherence and noise levels. When the objective involves optimizing for a lower energy solution, QAC techniques \cite{QAC1,QAC2,QAC3,QAC4,QAC5,QAC6,QAC7,QAC8,QAC9,QAC10,QAC11} may represent more appropriate approach to enhance overall performance.
The technique described in this work can be applied to tackle more intricate challenges, such as quantum simulations of exotic magnetic materials. Leveraging ZNE makes it feasible to estimate expectation values in situations where classical computation approaches become intractable.

\subsection*{Acknowledgment}

We would like to thank Gonzalo Alvarez, Jacek Dziarmaga, Daniel Lidar, Hidetoshi Nishimori,  and Marek Rams for fruitful discussions and comments on the manuscript. Work at UBC was supported by NSERC Alliance Quantum Program (Grant ALLRP-578555), CIFAR and the Canada First Research Excellence Fund, Quantum Materials and Future Technologies Program.  

\appendix

\section{Open quantum model}

In this appendix we use open quantum modeling to explain ZNE. We write the total Hamiltonian as 
\be
H = H_S + H_B + H_{\rm int}, \label{HSB}
\ee
where $H_S$ and $H_B$ are system and bath Hamiltonians, respectively. We write the interaction Hamiltonian in a general form as 
\be
 H_{\rm int} = - \sqrt{\gamma} \sum_\alpha Q^\alpha {\cal O}^\alpha \label{Hint}
\ee
where ${\cal O}^\alpha$ is an operator acting on the system and $Q^\alpha$ is the noise operator acting on the corresponding environment. The operators ${\cal O}^\alpha$ are typically Pauli matrices, e.g., $\sigma^z_\alpha$ for flux noise and $\sigma^y_\alpha$ for charge noise. For static noise, such as control error, $Q^\alpha$ can be considered as a constant random number.

In open quantum systems, the state of the system is described by the reduced density matrix $\rho(t)$. We choose the preferred basis to be the eigenstates $\ket{n}$ of the system Hamiltonian $H_S$ with eigenvalues $E_n$. 
The Bloch-Redfield equation for the reduced density matrix $\rho(t)$ is written as \cite{BR}
\ba
 \dot \rho_{nm} (t) = - i\omega_{nm}\ \rho_{nm} (t)
 - \gamma \sum_{k,l} R_{nmkl} \ \rho_{kl}(t) \label{BR}
\ea
where $\omega_{nm} = E_n-E_m$ and the Redfield tensor is defined as
\ba
 R_{nmkl} &=& \delta_{lm} \sum_r \Gamma^{(+)}_{nrrk} +
 \delta_{nk} \sum_r \Gamma^{(-)}_{lrrm} \nn
 && -\Gamma^{(+)}_{lmnk} - \Gamma^{(-)}_{lmnk}
 \label{Rlmnk}
\ea
with
\ba
 && \Gamma^{(+)}_{lmnk} = {1\over 2}\sum_{\alpha,\beta} {\cal O}^\alpha_{lm} {\cal O}^\beta_{nk}  S_{\alpha\beta}(-\omega_{nk}), \label{Gmp} \\
 && \Gamma^{(-)}_{lmnk} = {1\over 2}\sum_{\alpha,\beta} {\cal O}^\alpha_{lm} {\cal O}^\beta_{nk}  S_{\alpha\beta}(\omega_{lm}),
 \label{Gmm}
\ea
Here, ${\cal O}^\alpha_{nm} \equiv \bra{n} {\cal O}^\alpha \ket{m}$ and the noise spectral density is defined as \cite{Note3}
\ba
 S_{\alpha\beta} (\omega)=\int_{-\infty}^\infty dt \ e^{i\omega t} \langle Q^\alpha(t) Q^\beta(0)
 \rangle.
\ea
For the case of uncorrelated heat baths we have 
\be
S_{\alpha\beta} (\omega) = \delta_{\alpha\beta} S_\alpha (\omega).
\ee
For a time-dependent system Hamiltonian, the generalized Bloch-Redfield equation becomes \cite{Amin2009}
\ba
 \dot \rho_{nm} &=& -i\omega_{nm} \rho_{nm} - \sum_{kl}
 \left( \gamma R_{nmkl} +  M_{nmkl} \right) \rho_{kl}. \label{GBR}
\ea
where
\ba
 M_{nmkl} = -\delta_{nk} \langle l| \dot m \rangle -
 \delta_{ml} \langle \dot n|k \rangle \label{Mnmkl}.
\ea

In the energy basis, we write the density matrix as a linear vector with $2^{2N}$ elements:
\be
\hat \rho = \left[ \begin{array}{c}
     \rho_{11}  \\
     . \\
     . \\
     \rho_{nm} \\
     . \\
     . 
\end{array} \right]
\ee
The master equation becomes a matrix equation:
\be
{d \over dt} \hat \rho(t) = [-i \hat \omega(t) - \hat M(t) - \gamma \hat R(t) ] \hat \rho(t)
\ee
where
\be
\hat \omega(t) = \left[ \begin{array}{ccccccc}
    \omega_{11}(t) &  &  &  &  &  & \\
     & . &  &  &  &  & \\
     &  & . &  &  &  & \\
     &  &  & \omega_{nm}(t) &  &  & \\
     &  &  &  & . &  & \\
     &  &  &  &  & . & 
\end{array}
\right],
\ee
is a diagonal matrix made of energy differences and $\hat M$ and $\hat R$ are the basis-rotation and relaxation matrices corresponding to $M_{nmkl}$ and $R_{nmkl}$, respectively. A formal solution to this equation is
\be
\hat \rho(t) = \hat C_1 + {\cal T} e^{- \int_0^t [i \hat \omega(\tau) + \hat M(\tau) +  \gamma \hat R(\tau)] d\tau} \hat C_2. 
\ee
The time-ordering operator ${\cal T}$ is introduced to take care of non-commuting $\hat \gamma(t)$, $\hat M(t)$, and $\hat R(t)$ at different times. Changing the integration variable to $s=\tau/t_a$, we get
\be
\hat \rho(t_a) = \hat C_1 + {\cal T} e^{- t_a \int_0^1 [i \hat \omega(s) + \hat M(s) + \gamma \hat R(s)] ds} \hat C_2. 
\ee
The term $\gamma \hat R(s)$ in the exponent take care of all relaxation and dephasing processes during the evolution. In the limit of small $\gamma$ when those decay processes are not very strong it is possible to assume this term is small and expand the exponent. 
Using Taylor expansion, we can write
\be
\hat \rho(t_a) =  \hat \rho_0(t_a) + \sum_{\mu=1}^\infty  (\gamma t_a)^\mu \, \hat\rho_\mu(t_a),
\ee
where
\be
\hat \rho_0(t_a) = \hat C_1 + {\cal T} e^{- t_a \int_0^1 [i \hat \omega(s) + \hat M(s)] ds} \hat C_2 
\ee
is the noise-free contribution to the density matrix and
\be
\hat \rho_\mu(t_a) = \Bigg[{d^\mu \over \mu ! (t_a d \gamma)^\mu} {\cal T} e^{- t_a \int_0^1 [i \hat \omega(s) + \hat M(s) + \gamma \hat R(s)] ds} \bigg]_{\gamma = 0} \hat C_2, 
\ee
captures the effect of noise to order $\mu$ in perturbation.  Turning back the density matrices from vector form to matrix form we obtain \eqref{rhoexpansion}.

In some cases, only a small number of energy states get occupied during the annealing process. Therefore, we only need to incorporate relaxation and dephasing processes between those states. The experiment reported in Ref.~\cite{Dickson13} is one such case. The evolution is limited to the lowest two energy states between which the relaxation process is extremely slow. This allows expansion in powers of those relaxation rates only, although other decay processes could be much faster. As a result zero-T extrapolation worked for time scales millions of times longer than the expected single qubit decoherence time of the qubits.

\section{Energy-time rescaling}

The goal of this section is to introduce a relation between energy and time rescaling so that the critical dynamics remains unchanged. We first write the time-dependent Hamiltonian in a dimensionless form with one dimensionless parameter $\tau_Q$, which measures the speed of passing the critical point. The same quantity is used in some theory papers (see e.g. \cite{Nowak,Dziarmaga2022}), hence our formalism would allow direct comparison with these analytical results.

We start by writing the Hamiltonian \eqref{HS} as
\begin{equation} \label{e2}
 H(s) =   \mathcal J(s)   \bigg(\!\!- g(s) \sum_i  \sigma_i^x  + H_P \bigg),
\end{equation}
where
\be
g(s) = {\Gamma(s) \over \mathcal J(s)}
\ee
is the dimensionless transverse field. The corresponding Schrodinger equation reads
\be \label{seq}
i {d \over d t} \ket{\psi(t)} =   \mathcal J(s)   \bigg(\!\!- g(s) \sum_i  \sigma_i^x  + H_P \bigg)\ket{\psi(t)}.
\ee
Introducing a dimensionless time
\be
\tilde t = \int_0^t \mathcal J(t') dt' = t_a  \int_0^s \mathcal J(s') ds', \label{ttlde}
\ee
we can write \eqref{seq} in dimensionless form
\be \label{seq2}
i {d \over d \tilde t} \ket{\psi(\tilde t)} = \tilde H (\tilde t) \ket{\psi(\tilde t)}
\ee
where 
\be
\tilde H (\tilde t) =  - g(\tilde t) \sum_i  \sigma_i^x  + H_P
\ee
is the dimensionless Hamiltonian. Note that $\tilde t \in [0, \tilde t_a]$ with $\tilde t_a = t_a \int_0^1 \mathcal J(s) ds$ being the dimensionless annealing time.  

For fast annealing, only the speed of passing the critical point determines the critical dynamics. We define the critical point $s = s_c$ by
 \be \label{CPoint}
 \Gamma(s_c) /\mathcal J(s_c) = g_c
 \ee
where $g_c$ is the critical value of $g$. Using linear expansion near the critical point we write
\be \label{ttQ}
g(\tilde t) \approx g_0 - {\tilde t \over \tau_Q}
\ee
where $\tau_Q$ is the dimensionless quench time scale. It is possible to express Kibble-Zurek dynamics only in terms of $\tau_Q$. In other words, two systems with the same $\tau_Q$ are expected to yield the same statistics as long as only critical dynamics affect their evolution. For a slower evolution where the dynamics outside the critical region affect the results, details of the schedule become important. 
 
We expand the schedule near the critical point as
\ba
\Gamma(s) &=& \Gamma(s_c) +  \Gamma'(s_c) (s - s_c) 
\label{expn1}\\
\mathcal J(s) &=& \mathcal J(s_c) +  \mathcal J'(s_c) (s - s_c).
\label{expn2}
\ea
Expanding $g(s)$ near $s_c$, we obtain
\ba \label{e5}
g(s) &\approx&  g_c +  {\Gamma'(s_c)\mathcal J(s_c) -  \Gamma(s_c)\mathcal J'(s_c) \over \mathcal J^2(s_c)} (s - s_c) \nn
 &\approx&  g_c +  g_c \bigg[ {\Gamma'(s_c) \over  \Gamma(s_c)} - {\mathcal J'(s_c) \over \mathcal J(s_c)} \bigg] (s - s_c) \nn
 &\approx&   g_c \bigg[ {\Gamma'(s_c) \over  \Gamma(s_c)} - {\mathcal J'(s_c) \over \mathcal J(s_c)} \bigg] {t \over t_a} + const.
\ea
On the other hand
\be \label{e4}
\tilde t \approx \mathcal J(s_c) t + const.
\ee
Therefore
\be
g(\tilde t) \approx  g_c \bigg[ {\Gamma'(s_c) \over  \Gamma(s_c)} - {\mathcal J'(s_c) \over \mathcal J(s_c)} \bigg] {\tilde t \over  \mathcal J(s_c) t_a} + const.
\ee
Comparing  with \eqref{ttQ}, we obtain
\be \label{tauQ}
\tau_Q = {t_a \over t_Q},
\ee
where
\be \label{tQ}
t_Q = {g_c \over \mathcal J(s_c)} \bigg[ {\mathcal J'(s_c) \over \mathcal J(s_c)} - {\Gamma'(s_c) \over  \Gamma(s_c)} \bigg] 
\ee
is a time scale that only depends on the annealing schedule. For 1D spin chain, we have $g_c=1$ and the critical point is the point where $J(s_c) = \Gamma(s_c)$.  On the other hand, for 2D and 3D problems we have $g_c>1$ and therefore we expect the critical point to occur earlier in the schedule.

In order for the two systems to have the same critical dynamics, they need to have the same $\tau_Q$.  Equation \eqref{tauQ} requires that $t_a$ should be scaled the same way as $t_Q$ in \eqref{tQ}. If both $\Gamma(s)$ and ${\cal J}(s)$ are scaled the same way such that $H_S \to H_S/\kappa$, we obtain $t_Q \to \kappa t_Q$, therefore we need $\lambda = \kappa$, as expected for simple rescaling.  

For the more practical case of only rescaling  $H_P$, it is equivalent ${\cal J}(s) \to {\cal J}(s)/\kappa$ without changing $\Gamma(s)$. Obviously, the critical point will be shifted to a new point $s_c^\kappa$ which is the solution to $\Gamma(s_c^\kappa) = g_c \kappa {\cal J}(s_c^\kappa)$. Thus, one needs to calculate the new $t_Q^\kappa$, obtained from \eqref{tQ} at $s_c^\kappa$. The correct time rescaling coefficient will therefore be:
\be
\lambda(\kappa) = t_Q^\kappa/t_Q.
\ee
For case of the 1D transverse field Ising problem, we can write
\be
\lambda(\kappa) = { \mathcal J(s_c)^2 [ \mathcal J'(s_c^\kappa) - \Gamma'(s_c^\kappa)] \over \mathcal J(s_c^\kappa)^{2} [ \mathcal J'(s_c) - \Gamma'(s_c) ]}.
\ee

\section{Experimental details}

\begin{figure}
    \includegraphics[scale=0.8]{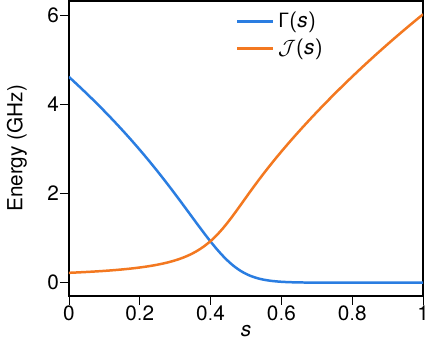}
    \caption{QPU annealing schedule.  $\Gamma(s)$ and $\mathcal J(s)$ are estimated using single-qubit measurements, then both are rescaled by an overall fit parameter $1/1.15$ to account for any deviation between single-qubit measurements and many-body flux qubit behavior. The y-axis units is GHz $= 10^{-9}$ Joules/$h$, where $h$ is the Planck constant. For equations with $\hbar = 1$ convention, these numbers should be multiplied by $2\pi$. }\label{fig:schedule}
\end{figure}

Quantum annealing experiments were performed using a prototype D-Wave Advantage2 processor, with 232 flux qubits coupled in a one-dimensional periodic chain.  At each temperature, coupling strength, and annealing time, calibration refinement was performed to homogenize coupler frustration, balance each qubit at zero magnetization, and synchronize annealing schedules.  These refinements are performed by tuning, respectively, individual couplings, flux-bias offsets, and anneal offsets as previously detailed in the supplementary materials of \cite{King22} and as described in \cite{Chern23} (the latter does not employ or describe anneal offset tuning, as it is only impactful for anneals faster than 100\,ns).

Average kink densities are measured over 20 programmings; 100 samples are taken in each programming.  Kink-kink correlations require more robust statistics, and are measured over 1000 programmings; 1000 samples are taken in each programming.  All error bars represent 95\% confidence intervals generated from bootstrap resampling of the 20 or 1000 programmings. 

Qubit temperatures are measured by the population slope for uncoupled qubits subjected to a varying longitudinal field.

Zero-noise extrapolation was performed using linear extrapolation in $T$ (Fig.~\ref{fig1}), quadratic extrapolation in $\lambda$ (Fig.~\ref{fig2}), and linear extrapolation in $\lambda$ (Fig.~\ref{fig3}) to zero.  Relative values of $\lambda$ were determined from an estimated annealing schedule shown in Fig.~\ref{fig:schedule}, which was derived from single-qubit measurements; rather than refining the single-qubit schedule with detailed many-body modeling, as in previous works \cite{King22,King23}, we simply introduced a single parameter and multiplied both $\Gamma(s)$ and $\mathcal J(s)$ by $1/1.15$ to align measured and simulated kink densities in the 5-10\,ns region.

\section{Details about DMRG simulations}

\begin{figure}[h]
\begin{center}
\includegraphics[scale=0.3]{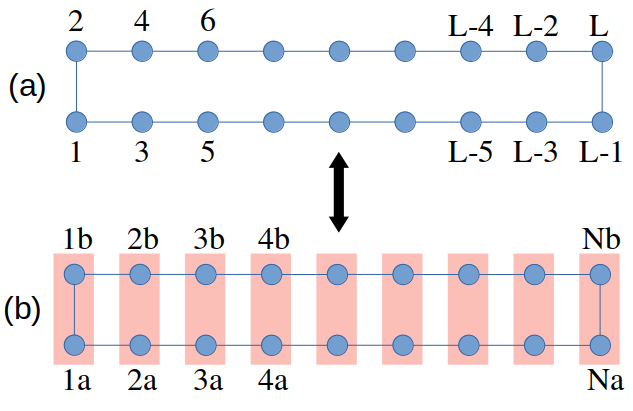}
\caption{\label{figA} (a) Site ordering for MPS simulations of a periodic chain with one-qubit local Hilbert space. (b) Site ordering with a two-qubit local Hilbert space.}
\end{center}
\end{figure}

For kink density as well as kink-kink correlations, we 
compared QA output against results from exact solution 
via Jordan-Wigner transformation and against DMRG, with 
dynamics simulated by the time-dependent variational 
principle (TDVP) with a two site 
update~\cite{Haegeman2011,Haegeman2016} using ITensor 
library~\cite{itensor}. We also compared the results 
against the $W_{\text{II}}$ method introduced in 
Ref.~\cite{Zaletel2015} using the TenPy 
library~\cite{TenPy} and the local Krylov 
method~\cite{Alvarez2011} (for details about this method 
in MPS language see Ref.~\cite{Paeckel2019}) implemented 
by the DMRG++ library~\cite{DMRGpp}. Overall, we found 
that TDVP emerges as the most efficient method, providing 
converged results using the largest time 
step ($dt$=0.01\,ns). 
For the lattice geometry, we adopted the same mapping from periodic to open chain used in Ref.~\cite{King22} as shown in Fig.~\ref{figA}a.
We found a much better performance of the numerical simulations when using a local Hilbert space of two qubits (see Fig.~\ref{figA}b) for the MPS wave-functions.
For the results shown in the main part of the manuscript, TDVP simulations were performed with a bond dimension $D=32$ and a time step of 0.01\,ns, with a maximum truncation error of $10^{-10}$.


\begin{thebibliography}{1}
\bibitem{Note} In general, different qubits may be coupled to different heat baths with different spectral densities, but these details don't affect our argument. See Appendix A for more details.
\end{thebibliography}
\end{document}